\documentclass[pdflatex,iicol, sn-mathphys-num]{sn-jnl}

\usepackage{graphicx}%
\usepackage{multirow}%
\usepackage{amsmath,amssymb,amsfonts}%
\usepackage{amsthm}%
\usepackage{mathrsfs}%
\usepackage[title]{appendix}%
\usepackage{xcolor}%
\usepackage{textcomp}%
\usepackage{manyfoot}%
\usepackage{booktabs}%
\usepackage{algorithm}%
\usepackage{algorithmicx}%
\usepackage{algpseudocode}%
\usepackage{listings}%

\usepackage{makecell}

\begin{document}

\title[Article Title]{Extreme black points in Born-Infeld electrodynamics}

\author[1]{\fnm{V.A.} \sur{Sokolov }}\email{sokolov.sev@inbox.ru}

\affil[1]{\orgdiv{Physics Department}, \orgname{Moscow State University}, \orgaddress{ \city{Moscow}, \postcode{119991}, \country{Russia}}}

\abstract{The article considers the space-time structure of a charged black hole in the nonlinear Born-Infeld electrodynamics. We are discussing a special state of such a black hole in the form of a "black point" with a doubly degenerate horizon, for which the pseudo-Riemannian spacetime has a timelike singularity, and the effective space-time for photons turns out to be everywhere regular. This property makes extreme black points an intermediate state between traditional and absolutely regular black holes.}

\keywords{regular black hole, Born-Infeld electrodynamics, effective metric, black hole shadow}

\maketitle

\section{Introduction}\label{sec1}

One of the fundamental problems of field theory associated with
the infinite value of the electromagnetic field energy of a point
charge attracts considerable interest and has several alternative
solutions. In quantum electrodynamics, the elimination
of divergence is ensured by using a renormalization scheme, 
which does not always lead to unambiguous results. 
Regularizing the field energy with higher derivatives in the
Lagrangian \cite{a3, a4} is also a solution to the problem. 
However, this approach increases the order of
the dynamical field equations and requires a priori information
to eliminate redundant solutions.

The modification of the electromagnetic field Lagrangian can
be realized without using higher derivatives of the field
variables. In this case, to solve the regularization problem, it is necessary
to resort to models of nonlinear vacuum electrodynamics, 
among which Born-Infeld electrodynamics \cite{a6} occupies a special place.
First constructed on the basis of heuristic ideas, this model was
reproduced on a D-brane in the low-energy limit of string theory \cite{a7}. 
Notable features of Born-Infeld electrodynamics are the absence
of vacuum birefringence~\cite{CH1_BI_BFfree2} and shock waves 
\cite{CH1_BI_SWfree2, CH1_BI_SWfree3}, while 
maintaining the dual symmetry inherent in Maxwell electrodynamics.

The nonlinear properties of Born-Infeld electrodynamics lead to
modification of the known exact analytical solutions describing
compact astrophysical objects, in particular, charged black holes. 
In Einstein-Maxwell theory, the metric of such a black hole is
represented by the Ressiner-Nordstr\"{o}m solution. 
In the transition to Born-Infeld electrodynamics, the solution
undergoes a substantial modification, 
resulting in the acquisition of novel properties, 
which will be discussed in more detail.

The metric function $g_{00}$ of an Einstein-Born-Infeld 
black hole with mass $M$ and electric charge $Q$ are
determined by the expression \cite{S2011}:
\begin{equation}
g_{00}=1-\frac{2M}{r}+\frac{2}{a^2 r}\int\limits^\infty_r
\big[\sqrt{\eta^4+a^2Q^2}-\eta^2\big]d\eta, \label{SolFinalQM}
\end{equation}
where $a$ is a parameter with the meaning of the inverse
electric field strength at the center of a point charge. 
The number of horizons for the metric \eqref{SolFinalQM} depends on the mass-to-charge ratio of the black hole \cite{S2016c}. Of considerable interest is the possibility of the existence of
a degenerate state in which the event horizon
has zero radius and coincides with the
singularity position. This state is known
for the logarithmic electrodynamics model \cite{CH1_BP_Soleng} and is called a "black point".

For the Einstein-Born-Infeld black hole, a black point state with a
doubly degenerate horizon becomes possible, similar in meaning
to the extreme state of the Reissner-Nordstrom black hole, 
which occurs at $|Q|=M$ (in the natural system of units). 
To realize the state of extreme black point, 
it is necessary that the equation $g_{00}=0$
have a second-order root at $r=0$, 
which leads to the values of the critical mass
and charge essential for the existence of this state:
\begin{equation}
M_{cr}=\frac{1}{3\Gamma^2(3/4)}\Big(\frac{\pi}{2}\Big)^{3/2}a\simeq 0.437 a, \quad 
|Q_{cr}|=a/2.\label{Mcr_IntRel}
\end{equation}
A quantitative estimate of the critical mass for the accepted  value \cite{a6}  of 
the Born-Infeld parameter
$a=2374 \cdot M_{\odot}$
leads to $M_{cr}=1037.8 M_{\odot}$, 
which attracts attention to the study
of intermediate-mass black holes \cite{CH3_BH_Marel2004, CH3_BH_Farell2009}.

The unusual state of the extreme black point
gives considerable interest to the study of the question about the peculiarities of light propagation near it, 
which will be the main purpose of this article.

\section{Isotropic geodesic lines near extreme black point and the effective metric}
The description of electromagnetic wave propagation in the field of
compact astrophysical objects is a non-trivial problem, especially
in essentially nonlinear field theories such as Born-Infeld electrodynamics. 
Significant progress in this issue has been achieved by applying
the geometrized approach or the concept of natural geometry, which
allows one to reduce the equation of isotropic geodesic rays to the form of the electromagnetic wave front equation in an effective
spacetime with a metric tensor $G_{ik}$ whose components depend on the metric tensor of the pseudo-Riemannian spacetime $g_{ik}$ and the tensor of the background electromagnetic field $F_{ik}$ 
in which the electromagnetic wave propagates.
For Born-Infeld electrodynamics, the components of the metric tensor
of the effective spacetime do not depend
on the wave polarization and have the form:
\begin{equation}\label{EffectiveMetr_QM}
G_{kn}=\frac{g_{kn}-a^2F_{kn}^{(2)}}{
1-\dfrac{a^2}{2}J_2-\dfrac{a^4}{4}J_4+\dfrac{a^4}{8}J_2^2},
\end{equation}
where $J_2=F_{ik}F^{ki}$, $J_{4}=F_{ik}F^{kl}F_{lm}F^{mi}$ are
the invariants of the electromagnetic field tensor and $F_{kn}^{(2)}=F_{km}F^{m\cdot}_{\cdot n}$.
In the particular case of a charged Einstein-Born-Infeld black
hole \eqref{SolFinalQM} the non-trivial components of the
effective metric will be as follows: 
\begin{align}
G_{00}&=g_{00}, \quad G_{11}=g_{11}=-1/g_{00}, \label{EffMetr_Comp_QM}
\\
G_{22}&=\frac{ r^4+a^2Q^2}{r^4}g_{22}=-\Big(r^2+\frac{a^2Q^2}{r^2}\Big),
\nonumber \\
G_{33}&=\frac{ r^4+a^2Q^2}{r^4}g_{33}=G_{22}\sin^2\theta. \nonumber
\end{align}
The difference of pseudo-Riemannian and effective metrics
can testify about the essentially different properties
of motion for photons and massive particles near a charged black hole. To clarify this question, let us consider the properties of singularity for each of the two types of spacetime.
For this purpose we calculate the scalar curvature $R$ for the
pseudo-Riemannian metric \eqref{SolFinalQM} and the scalar curvature ${\cal R}$ for the
effective metric \eqref{EffMetr_Comp_QM}, and also we determine their
asymptotic behavior near the center of a black hole  
\begin{gather}
R=\frac{4Q^2}{r^2\sqrt{r^4+a^2Q^2}}
\Bigg[\frac{r^2-\sqrt{r^4+a^2Q^2}}{r^2+\sqrt{r^4+a^2Q^2}}\Bigg]\Bigg|_{r\to 0}\simeq \nonumber \\
\simeq \frac{8}{a^2}-\frac{4|Q|}{a r^2}-\frac{6 r^2}{a^3 |Q|}+{\cal O}(r^3),  \\
{\cal R}=\frac{16a^2Q^2(3r^4+2a^2Q^2)}{r^3(r^4+a^2Q^2)^2} 
\Bigg[Q^2\int\limits_r^\infty \frac{d\eta}{\eta^2+\sqrt{\eta^4+a^2Q^2}}
\nonumber \\
-M\Bigg]+\frac{2(4r^{10}+9r^4Q^2a^4+5a^6Q^4-4r^2a^4Q^4)}{a^2r^2(r^4+a^2Q^2)^2} \nonumber \\
+\frac{4(3Q^2a^2-2r^4)}{a^2r^2\sqrt{r^4+a^2Q^2}}\Bigg|_{r\to 0} \simeq \nonumber \\
\simeq \frac{32(M_{cr}-M)}{r^3}+\frac{10}{r^2}\Big(1-\frac{2|Q|}{a}\Big)
+\frac{8}{3a^2}- \nonumber \\ 
-\frac{16(M_{cr}-M)r}{a^2Q^2}+\frac{2(|Q|-a)r^2}{Q^2a^3}+{\cal O}(r^3).
\end{gather}
In the general case, both expressions are singular in the center of the black hole,
but for an extreme black point with mass and charge parameters \eqref{Mcr_IntRel} 
the scalar curvature ${\cal R}$ becomes regular everywhere, 
while the singularity for $R$ is preserved. 

The assertion that there is no singularity in the effective space-time for a black point cannot be made solely on the basis of the expression for the scalar curvature. As a rule, three
invariants of the curvature tensor are used to check the
regularity of static solutions of the Einstein equations:
scalar curvature, quadratic invariant of the Ricci tensor
and the Kretschmann scalar. However, this set of invariants
is not always complete and the only possible one. 
A special place is occupied by a set of invariants,
proposed by Carminati and McLenaghan in the
paper \cite{CH3_BH_CarminatiMcLenaghan}. 
When introducing this set, the authors sought
to ensure two conditions: the set should be constructed
from invariants with the lowest degree of tensor quantities, 
and it should contain the minimum number of independent
invariants in each of the spaces according to the
Petrov-Segre classification.  
The set of Carminati-McLenahan invariants
includes the following expressions:     
\begin{align}
R=&R_{m \ }^{\ m},  \nonumber \\
R_1=&\frac{1}{4}S_{a \ }^{\ b}S_{b \ }^{\ a}, \nonumber \\
R_2=&-\frac{1}{8}S_{a\ }^{\ b}S_{b\ }^{\ c}S_{c \ }^{ \ a},\nonumber \\
R_3=&\frac{1}{16}S_{a\ }^{\ b}S_{b\ }^{\ c}S_{c \ }^{ \ d}S_{d \ }^{ \ a},\nonumber \\
M_1=&\frac{1}{8}S^{ab}S^{cd}(C_{acdb}+i{}^{*}C_{acdb}), \nonumber \\
M_2=&\frac{1}{16}S^{bc}S_{ef}(C_{abcd}C^{aefd}+{}^{*}C_{abcd}{}^{*}C^{aefd})+ \nonumber \\
&+\frac{1}{8}i S^{bc}S_{ef}{}^{*}C_{acdb}C^{aefd}, \nonumber \\
M_3=&\frac{1}{16}S^{bc}S_{ef}(C_{abcd}C^{aefd}+{}^{*}C_{abcd}{}^{*}C^{aefd}),\nonumber  \\ 
M_4=&-\frac{1}{32}S^{ag}S^{ef}S^{c\ }_{\ d}(C_{ac}^{\ \ db}C^{befg}+{}^{*}C_{ac}^{\ \ db} \ {}^{*}C^{befg}),
\nonumber \\
M_5=&\frac{1}{32}S^{cd}S^{ef}(C^{aghb}+i{}^{*}C^{aghb})\times \nonumber \\
&\times(C_{acdb}C_{gefh}+{}^{*}C_{acdb}{}^{*}C_{gefh}), 
\nonumber \\
W_1=&\frac{1}{8}(C_{abcd}+i{}^{*}C_{abcd})C^{abcd}, 
\nonumber \\
W_2=&-\frac{1}{16}(C_{ab}^{\ \ cd}+i{}^{*}C_{ab}^{\ \ cd})C_{cd}^{\ \ ef}C_{fe}^{\ \ ab},
\end{align}
where we use the notations $S_{ab}=R_{ab}-R g_{ab}/4$ for the
deviator of the Ricci tensor, $C_{abcd}$ for the components of the Weyl tensor
and the asterisk denotes the dual conjugation. 
It was shown in \cite{CH3_BH_Zakhary} that this set of invariants
is complete for all known types of spaces associated with the
electrovacuum solutions in Einstein-Maxwell theory, as well as
for a number of spaces in which other sets of invariants
are incomplete. In spite of the fact that to date
there is no proof of the completeness of this set for all 90 types
of spaces according to the Petrov-Segre classification, we will use it to find out 
the regularity of black point space-time 
in Einstein-Born-Infeld theory. 
The results of computing the Carminati-McLenahan invariants
for an extreme black point at $M=M_{cr}$ and $|Q|=a/2$ are
summarized in Table~\ref{InvarKM_QM}. 
The expressions are presented as segments of a series expansion
near the center of the black hole and are calculated for both
the pseudo-Riemannian metric $g_{ik}$ and the effective spacetime metric $G_{ik}$. 

\begin{table*}[htb]
\caption{Asymptotic of the Carminati-McLenahan invariants near the extreme black point center.} \label{InvarKM_QM}
\centering
\begin{tabular}{@{}lll@{}}
\toprule
The invariant  notation & Expression for the metric $G_{ik}$&  Expression  for the metric $g_{ik}$ \\[2ex]
\midrule
 $R$ & $\dfrac{8}{3a^2}-\dfrac{4r^2}{a^4}+{\cal O}(r^3)$ &
 $-\dfrac{2}{r^2}+\dfrac{8}{a^2} -\dfrac{12 r^2}{a^4} +{\cal O}(r^3)$ \\[1.5ex] 
\midrule
 $R_1$ & $\dfrac{8}{9a^4}+{\cal O}(r^3)$ & $\dfrac{1}{4r^4}+{\cal O}(r^3)$ \\[1.5ex]  
\midrule
$R_2$ & $-\dfrac{8r^2}{3a^8}+{\cal O}(r^3)$ & ${\cal O}(r^3)$ \\[1.5ex]  
\midrule
 $R_3$ & $\dfrac{32}{81a^8}+{\cal O}(r^3)$ & $\dfrac{1}{64r^8}+{\cal O}(r^3)$ \\[1.5ex]  
\midrule
 $M_1$ & $\dfrac{16}{81a^6}-\dfrac{16r^2}{27a^8}+{\cal O}(r^3)$ & $\dfrac{1}{12r^6}+{\cal O}(r^3)$ \\[1.5ex]  
\midrule
$M_2=M_3$ & $\dfrac{32}{729a^8}-\dfrac{64r^2}{243a^{10}}+{\cal O}(r^3)$ & 
$\dfrac{1}{36r^8}+{\cal O}(r^3)$ \\[1.5ex]   
\midrule
 $M_4$ & $\dfrac{32r^2}{243a^{12}}+{\cal O}(r^3)$ & ${\cal O}(r^3)$ \\[1.5ex]    
\midrule
 $M_5$ & $\dfrac{64}{6561a^{10}}-\dfrac{64r^2}{729a^{12}}+{\cal O}(r^3)$ & 
$\dfrac{1}{108r^{10}}+{\cal O}(r^3)$ \\[1.5ex]    
\midrule
 $W_1$ & $\dfrac{8}{27a^4}-\dfrac{16r^2}{9a^6}+{\cal O}(r^3)$ & 
$\dfrac{1}{6r^4}+{\cal O}(r^3)$ \\[1.5ex] 
\midrule
$W_2$ & $-\dfrac{16}{243a^6}+\dfrac{16r^2}{27a^8}+{\cal O}(r^3)$ & 
$\dfrac{1}{36r^6}+{\cal O}(r^3)$ \\[1.5ex] 
\botrule
\end{tabular} 

\end{table*}
The expressions for all invariants of the effective spacetime are regular in the center, which confirms the earlier assumption about the regularity of this spacetime. This result illustrates a new, quite unusual, property of Einstein-Born-Infeld black holes, whose spacetime can
have singularity for massive particles and, at the same time, be regular for photons.  

Let us study the peculiarities of the motion of
photons in space with the effective metric \eqref{EffMetr_Comp_QM}
in more detail, starting by calculating
the properties of the shadow of such a black hole.
To write down the photon trajectory equation, it is convenient
to use the inverse radial coordinate $u=1/r$, as well as
the notation for the aiming parameter $b$:
\begin{gather}
\Big(\frac{du}{d\varphi}\Big)^2=u^4G_{22}\Big[\frac{G_{22}}{b^2}+G_{00}\Big]= \label{TrajectEQ_QM} \\ 
=(1+a^2Q^2u^2)
\Big[\frac{1+a^2Q^2u^4}{b^2}-u^2g_{00}(u)\Big]=
\Psi(u). \nonumber
\end{gather}
The solutions of equation $\Psi(u_c)=0$, corresponding to
the zeros of the second order, determine the radii of circular orbits, 
which have the meaning of limit cycles. 
The set of such orbits forms a photon sphere, 
which is essential for the calculation of the black hole shadow.
For the Reissner-Nordstr\"{o}m black hole in Einstein-Maxwell theory, the radius of the photon sphere decreases monotonically with increasing black hole charge and
takes the minimum value $r_c=1/u_c=6M$ for the extreme black hole $|Q|=M$.

In Einstein-Born-Infeld theory,
the expression \eqref{TrajectEQ_QM} leads to a
transcendental equation for the radius of the photon sphere:
\begin{gather}
(1-a^2Q^2u_c^4)+\frac{Q^2u_c^2(1+a^2Q^2u_c^4)}{1+\sqrt{1+a^2Q^2u_c^4}}+ \label{COrbitEQ_QM} \\
+(3-a^2Q^2u_c^4)\Bigg[{Q^2}\int\limits_0^{u_c}
\frac{d\xi}{1+\sqrt{1+a^2Q^2\xi^4}}-M\Bigg]u_c=0, \nonumber
\end{gather}
corresponding to the analogous equation in Reissner-Nordstr\"{o}m spacetime at $a\to 0$.

The angular size of the shadow, measured by an observer at a
point with radial coordinate $r=R$, 
is calculated as the angle between the light ray
that touched the surface of the photon sphere and
arrived at the observation point and the radial direction.

\begin{figure}[ht]
\begin{center}
\includegraphics[width=0.5\textwidth]{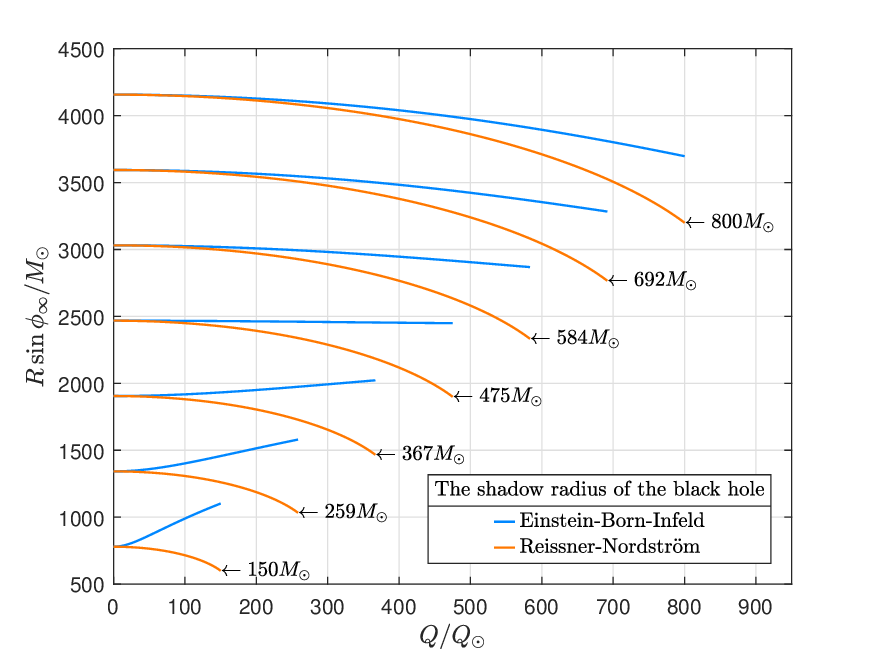}
\caption{The dependence of the shadow radius of the black hole on its charge 
for masses from $150 M_{\odot}$ to $800 M_{\odot}$.  
The charge is normalized to the maximum value of the charge allowed 
for a black hole of solar mass in Reissner-Nordstr\"{o}m spacetime. 
}\label{ShadowR_QM}
\end{center}
\end{figure}

After a simple transformation that takes into account
the photon trajectory equation \eqref{COrbitEQ_QM}, 
the expression for the angular size of the shadow can
be expressed in the form convenient for analysis:
\begin{gather}
\sin^2\phi=\Bigg[{1+\dfrac{1}{G_{00}|G_{22}|}
\Bigg(\dfrac{d r}{d\varphi}\Bigg)^2\Bigg]^{-1}\Bigg|_{r=R}}=\\
=\frac{G_{00}(R)}{|G_{22}(R)|}
\frac{|G_{22}(r_c)|}{G_{00}(r_c)}
=\frac{r_c^2+{a^2Q^2}/{r_c^2}}{R^2+{a^2Q^2}/{R^2}}
\ \Bigg(\frac{g_{00}(R)}{g_{00}(r_c)}\Bigg). \nonumber
\end{gather}
The expression for the geometric size of the shadow for a
distant observer $R\to \infty$ and an
asymptotically flat metric $g_{00}(R)\to 1$ takes the following form
\begin{equation}
R^2\sin^2\phi_\infty=\Big(1+\frac{a^2Q^2}{r_c^4}\Big)
\frac{r_c^2}{g_{00}(r_c)},
\end{equation}
and differs from the analogous expression in Einstein-Maxwell theory
by the multiplier in parentheses as well as by the
necessity to use the equation \eqref{COrbitEQ_QM} 
for the calculation of the radius of the photon sphere.
Figure \ref{ShadowR_QM} shows how the size of the black hole
shadow changes with its mass. 

For each mass, we varied the charge to its maximum limit in the Reissner-Nordstr\"{o}m solution. 
In the Einstein-Born-Infeld model, the maximum allowable charge for a given mass is slightly larger, but this is impossible to compare to the Reissner-Nordstr\"{o}m solution.
So, we excluded these values of the charge from the comparison.
The radius of the shadow decreases with increasing charge,
similarly to the radius of the photon sphere in the Reissner-Nordstr\"{o}m spacetime. 
However, at a black hole mass smaller than $M\approx 467 M\odot$ the shadow radius increases with charge.
This behavior is not typical for known types of black hole. 
It is caused by nonlinear features of Born-Infeld electrodynamics and can be
used as a signature of this theory in the analysis of observational data.

Finally, we note that in the case of the extreme black
point, at $M=M_{cr}$ and $|Q|=a/2$, the radius of
the shadow is related to the Born-Infeld parameter by a linear relationship $R\sin\phi_\infty \simeq 1.827 a$.

\section{Conclusion}
In this paper we have considered the peculiarities of light
propagation near a charged black hole in Born-Infeld electrodynamics, 
which allowed us to establish the possibility of existence of extreme
black points, surrounded by a twice degenerate horizon, 
with a new unusual property -- regularity of the invariants of
the curvature tensor of the photon effective spacetime, while
preserving the singularity of the similar invariants for the curvature
tensor of the pseudo-Riemannian spacetime.
This property distinguishes extreme black points, making them intermediate 
between  Ressiner-Nordstr\"{o}m black holes, whose spacetime for
massive and massless particles has a singularity at the center,
and fully regular black holes, for instance, \cite{CH1_REGBH_Ayon}.
Moreover, the leading terms in the expansions of the invariants
of the effective metric contain negative degrees of the Born-Infeld parameter $a$. 
This leads to the reconstruction of the singularity
in the limit of Maxwell's electrodynamics $a\to 0$.

Born-Infeld electrodynamics model was chosen for illustration because it is one of the most developed and well-studied. It is expected that a similar property of extreme black points will appear in other models of nonlinear electrodynamics with regularizing properties for point sources; in this respect, 
the model of rational electrodynamics \cite{CH1_Rational_Kruglov} seems very promising.

The size of the shadow of an extreme black point depends linearly
on the Born-Infeld parameter and, for its accepted value, 
reaches the range $R\sin\phi_\infty \simeq 4337 M_{\odot}$, in natural units.
Unfortunately, this value is at least three orders of magnitude
smaller than the limit on the size of the black hole's shadow
available for direct observation with modern instruments such
as the Event Horizon Telescope \cite{EHT2022}.  
However, the development of indirect detection methods for intermediate-mass black holes
\cite{InterMassBlackHoles_Review}, 
as well as the discovery of fairly realistic scenarios
for their charge accumulation \cite{CH3_PR_Ray2023}, allows us to
expect the appearance of new astrophysical observational data, 
which can clarify the status of vacuum nonlinear electrodynamics models.

The recent discovery of a compact astrophysical object in the galaxy NGC 4945, called "Punctum" \cite{shablovinskaia2025} seems extremely promising.
Such objects may possibly be associated with black points, and the high degree of polarization of their radiation can be a consequence of the vacuum birefringence effect inherent in a number of models of nonlinear vacuum electrodynamics (in contrast to the Born-Infeld model). The discovery of Punctum-type objects are of fundamental importance for the study of new non-perturbative effects in extremely strong electromagnetic and gravitational fields.

\section*{Acknowledgments}
I would like to express my sincere gratitude to my
colleagues K.A. Sveshnikov and D.A. Slavnov, whose memories will always be inspiring.
The study was conducted under the state assignment of Lomonosov Moscow State University.

\bibliography{BIRegular_Bib}

\end{document}